\documentstyle[aps,pre,preprint,floats]{revtex}
\tightenlines
\begin{document}
\input psfig.sty
\renewcommand{\H}{{\cal H}}
\newcommand{\N}{{\cal N}}
\newcommand{\Z}{{\cal Z}}
\newcommand{\D}{{\cal D}}
\newcommand{\k}{{\vec k}}
\voffset=0.5in
\title{\bf  
A generalized spherical version of the Blume-Emery-Griffiths model
with ferromagnetic and antiferromagnetic interactions.}
\author{S.E. Savel'ev$^a$ and Guillermo Ram\'\i rez-Santiago$^b$ }
\address{$(a)$ All-Russian Electrical Engineering Institute, Moscow,
Russia,
and Institute of Materials Science, University of Tsukuba, 1-1-1 Tennodai,
Tsukuba 305-8573, Japan,}
\address{$(b)$ Instituto de Fisica,
Universidad Nacional Aut\'onoma de M\'exico,
PO Box 20-364 M\'exico, 01000,
D.F. MEXICO}
\draft
\maketitle
\begin{abstract}
We have investigated analitycally the phase diagram of a generalized 
spherical version of the Blume-Emery-Griffiths model that includes 
ferromagnetic or antiferromagnetic spin interactions as well as quadrupole 
interactions in zero and nonzero magnetic field.
We show that in three dimensions and zero magnetic field  a regular
paramagnetic-ferromagnetic (PM-FM) or a paramagnetic-antiferromagnetic 
(PM-AFM) phase transition occurs whenever the magnetic spin interactions
dominate over  the quadrupole interactions. However, when spin and quadrupole
interactions  are important, there appears a reentrant FM-PM or AFM-PM  
phase transition at low temperatures, in addition to the regular PM-FM
or PM-AFM phase transitions.
On the other hand, in a nonzero homogeneous external magnetic field $H$,
we find no evidence of a transition to the state with spontaneous
magnetization for FM interactions in three dimensions.  
Nonethelesss, for AFM interactions we do get a scenario similar to that
described above for zero external magnetic field, except that the critical
temperatures  are now functions of $H$. We also find two critical field 
values, $H_{c1}$, at which the reentrance phenomenon dissapears and
$H_{c2}$ ($H_{c1}\approx 0.5H_{c2}$), above which the PM-AFM transition
temperature vanishes.
\end{abstract}
\vfill

\section{\bf Introduction}
Continuous spins models play an important role as approximate statistical
mechanical models for several physical systems. In particular, different
versions of the classical spherical model (CSM) introduced by Berlin and
Kac can be solved exactly for both, short\cite{berlin-kac} and long 
range \cite{joyce66,domb92} interactions. The spherical model (SM)
was originally introduced to study ferromagnetic systems in different
spatial and spin dimensions\cite{berlin-kac,domb92}. Unlike the mean-field 
approximation, in the case of spin dimensionality higher than one,
the spherical model rightly yielded no ferromagnetic transition
as had been observed experimentally.
Moreover, its equivalence to a spin system with infinity spin dimensions
has been completly established by Stanley\cite{stanley68}. Because of this, 
different
versions of the SM have been recently used to investigate the physics
of systems which hamiltonian can be maped out onto that of the spherical
model\cite{vojta96,others,kopec,aps99}.
The basic idea is the compactification and replacing
of too many restrictions by one spherical like  condition.
Thus, studying the phase transitions and the phase diagram of a 
spin ferromagnet or antiferromagnet using a particular
version of the SM is of physical relevance.  

In this paper we investigate the phase behavior of a generalized spherical
version of the Blume-Emery-Griffiths (BEG) model\cite{beg}. This is obtained 
from the discrete three states spin one ferromagnet or antiferromagnet and, as
in the original BEG model,  we include magnetic spin interactions as well as 
quadrupole  interactions. The BEG model  was introduced to understand 
qualitatively the phase 
separation, and the $\lambda$ transition of $^{3}$He-$^4$He mixtures.
It has also been applied to study  magnetic systems with competing
interactions\cite{jensen87} or to describe the qualitative phase behavior
of a microemulsion\cite{schick87}.

By introducing a transformation that maps out the original BEG model hamiltonian
onto that of the SM  we are able to investigate the conditions under which
phase transitions may occur. Proceeding as in the investigation 
of the phase transitions of the CSM we study the properties of the
saddle points of the integrand of the partition function. 

The main findings of this paper are the following:
(1) There is no phase transition at all in $d=1,2$ spatial dimensions for
both, the FM and AFM systems.
This result agrees well with that obtained for the CSM.
(2) For $d=3$  dimensions and zero external magnetic field, we obtain the
usual paramagnetic-ferromagnetic(antiferromagnetic) [PM-FM(AFM)]
phase transition  if the FM (AFM) interactions dominate over the quadrupole
interactions. On the opposite side, when the  quadrupole interactions
dominate over the magnetic exchange couplings no phase transition occurs.
This result is consistent with the limit of zero magnetic exchange
interactions. However, in the intermediate regime, when both,
quadrupole  and exchange interactions are relevant, there appears at low 
temperatures, a novel and intriguing reentrant phase transition, in addition
to the regular high temperature one. That is, the
FM or AFM  ordering takes place at intermediate
temperatures, since at high and low temperatures a PM phase sets in. 
(3) For a FM system in a nonzero external uniform magnetic field and 
$d=3$ spatial dimensions, the transition to a spontaneously magnetized
state is destroyed, so no phase transition occurs.
Nonetheless, for AFM exchange interactions in a uniform magnetic field
we find a scenario similar to that described in (2) for $H=0$, except that now
the critical temperatures are now functions of $H$. 
%
%That is,
%when the magnetic spin interactions dominate one gets the  usual
%PM-AFM phase transition at high  temperatures and, if both quadrupole  and
%magnetic spin interactions are relevant, a reentrant AFM-PM phase 
%transition appears at low 
%temperatures. On the other hand, when the quadrupole interactions
%are big enough, they supress any AFM ordering and no phase transition
%takes place. 
%
More importantly,  
the domain of parameters where the reentrant phase transition happens
narrows as the external magnetic field increases 
and, at the end, the transition completely disappears above a
certain critical value $H_{c1}$. Increasing further
the external magnetic field yields  a second critical field value 
$H_{c2}\approx 2H_{c1}$, above which the AFM phase is supressed.

The layout of the remainder of this paper is the following:
in section \ref{model} we
introduce the  BEG model hamiltonian and map it onto a generalized version 
of the spherical model. Then, we investigate the general
conditions under which phase transitions may take place.
In section \ref{short-range}  we consider the case of
short range interactions and study
the different scenarios that may occur in the system.
A novel global spherical condition that is fully equivalent to the SM
and valid even for low spin dimensionality is introduced in section
\ref{sec-global}.
Finally, in section \ref{conclusions} we
present a summary of our findings and the conclusions.

\section{\bf The model}
\label{model}
Let us  start with a BEG type model hamiltonian  with spins $S_{\vec R}=1$ 
located at each site of a $d$-dimensional rectangular lattice. Thus,
each spin has three different states namely,  $S^{\rm z}_{\vec R}=0, \pm 1$.
It is natural to assume that the spins interact with each other through
the pair potential $V(\vec R-\vec R')$ and with an effective external magnetic
field $H=M_0H_0$, where $H_0$ is a ``true external field" and $M_0$ is 
the magnetic moment produced by the states whith $S^{\rm z}_{\vec R}=\pm 1$.
According to these statements the Hamiltonian that defines the model is:
\begin{equation}
\H=E\sum_{\vec R}S_{\vec R}^2+\sum_{\vec R, \vec R'}
V\left({\vec R}-{\vec R'}\right)S_{\vec R}S_{\vec R'}
+H\sum_{\vec R}S_{\vec R},
\label{ham0}
\end{equation}
where the second sum may run over all nearest neighbors or over
all pairs of spins.
The first term in $\H$ has been introduced in order to have the
possibility of quadrupolar ordering, as suggested in reference \cite{beg},
in addition to the magnetic ordering associated with the second term.
In absence of an external magnetic field the competition between these two
types of ordering is measured by the parameter $E$, which can be interpreted as
the energy difference between the states with $S^{\rm z}_{\vec R}=\pm 1$ and
$S^{\rm z}_{\vec R}=0$.
With the aim at mapping this  model hamiltonian onto that of the 
spherical model, we introduce the Ising-like variables $a_{\vec R}=\pm 1$ and
$b_{\vec R}= \pm 1$ that are related to the original spin variables by,
\begin{equation}
S_{\vec R}=\frac{1}{2}\left(a_{\vec R}+b_{\vec R}\right).
\end{equation}
Under this transformation the state $S_{\vec R}=0$ has been double weighted
since it can now be obtained from the
states $a_{\vec R}=-b_{\vec R}= 1$ and $a_{\vec R}=-b_{\vec R}=-1$.
To compensate this {\it ``degeneration"} we introduce, in an {\it ad hoc} 
manner, the additional entropy term, $\H_{\cal S}=-
{{1}\over {4}}T\ln 2\sum_{\vec R}S^{2}_{\vec R}$, in units in
which the Boltzmann constant  $k_{B}$ is equal to unity.
In terms of these Ising-like new variables and including the entropy term
the hamiltonian $\H$ becomes:
\begin{eqnarray}
\H &=& \frac{1}{4}(E-T\ln 2) \sum_{\vec R}\left(a_{\vec R}+b_{\vec R}\right)^2
+\frac{1}{4}\sum_{\vec R, \vec R'}V\left({\vec R}-{\vec R'}\right)
\left(a_{\vec R}+b_{\vec R}\right)
\left(a_{\vec R'}+b_{\vec R'}\right)\nonumber\\
&+& \frac{H}{2}\sum_{\vec R}\left(a_{\vec R}+b_{\vec R}\right).
\label{ham1}
\end{eqnarray}
To be able to study analytically the properties of the model we have to make
contact with the classical spherical model \cite{berlin-kac,domb92}.
So, we replace the discrete
Ising-variables by continuous ones, that is, we let the ``new" spin variables
to vary in the interval,  $-\infty < a_{\vec R},
b_{\vec R} < +\infty $, obeying the spherical conditions:
\begin{equation}
\sum_{\vec R}a_{\vec R}^2=\N, \ \ {\rm and} \ \ \sum_{\vec R}b_{\vec R}^2=\N,
\label{spher1}
\end{equation}
where $\N$ is the total number of spins in the system. Hence,
the hamiltonian, Eqn.~(\ref{ham1}) together with the last two restrictions
can be considered as the {\it ``generalized spherical version"} 
of the BEG hamiltonian, Eqn.~(\ref{ham0}).
The reason of this name is that we have included
the additional entropy term that depends explicitly on temperature in the
quadrupole interaction part of the Hamiltonian. This term
does not appear in the original BEG model and will play a key role in the
results we describe in what follows.

Taking into account these spherical conditions on the ``new" Ising-like
variables, one can write  the partition function as follows:
\begin{equation}
\Z=\left\{\prod_{\{\vec R\}}\int_{-\infty}^{\infty}
da_{\vec R} \int_{-\infty}^{\infty} db_{\vec R}\right\}
\exp{\left(-\frac{\H}{T}\right)}\delta\left(\N-\sum_{\vec R}a^2_{\vec R}\right)
\delta\left(\N-\sum_{\vec R}b^2_{\vec R}\right).
\end{equation}

It is important  to point out that the transformation from discrete
Ising spin variables to the continuous spin ones is not
quite correct since there are ``wrong" configurations in
which there are spin variables that may take values different from
$\pm 1$. Nonetheless, such ``slight incorrect transformation" allows
one to solve analitycally two
classical problems: the 3D Ising Model\cite{berlin-kac,domb92},
and more importantly, the
3D Heisenberg model\cite{stanley68}, in addition to other more recent
contemporary physical problems\cite{vojta96,others,kopec,aps99}.
In section IV
we address this shortcoming and propose an absolutely correct
global spherical condition that is even valid for systems of low spin 
dimensionality, in particular for $\vec S_{\vec R} = 1$.
For the time being we proceed as in the CSM, that is, we use the
integral representation of $\delta$-function to rewrite the partition function
as:
\begin{equation}
\Z=\int\frac{ d s_1 d s_2}{(2\pi i)^2}\exp{\Bigl(\N(s_1+s_2)\Bigr)}
{\tilde \Z}\left(s_1,s_2\right),
\label{z-tot}
\end{equation}
where
\begin{equation}
{\tilde \Z}=\left\{\prod_{\vec R}\int_{-\infty}^{\infty} d a_{\vec R}
\int_{-\infty}^{\infty} d b_{\vec R}\right\}
\exp{\left(-\frac{{\tilde \H}(s_1, s_2)}{T}\right)},
\label{tz}
\end{equation}
and
\begin{equation}
{\tilde \H}=Ts_1\sum_{\vec R}a_{\vec R}^2+Ts_2\sum_{\vec R}b_{\vec R}^2+\H,
\label{th}
\end{equation}
where $\H$ is given in Eqn.~(\ref{ham0}), with the $\vec S_{\vec R}$
replaced by $\frac{1}{2}(\vec a_{\vec R} + \vec b_{\vec R})$.
In the framework of the CSM the effective hamiltonian $\tilde \H$ is  a
non-diagonal symmetric bilinear form in the new Ising-like variables.
To be able to integrate Eqn.~(\ref{tz}) one has to diagonalize
$\tilde \H$ by means of a  Fourier transformation of the spin variables.
In doing so one gets,
\begin{equation}
\Z=\pi^\N\int\frac{ds_1 ds_2}{(2\pi i)^2}\exp{\Bigl(\N\gamma(s_1, s_2)\Bigr)},
\label{int-z}
\end{equation}
where
\begin{equation}
\gamma\left(s_1, s_2\right)=s_1+s_2-\frac{1}{2\N}\sum_{\vec q}
\ln\left( s_1s_2+\left(s_1+s_2\right)\frac{\alpha (\vec q)}
{T}\right)
+\frac{H^2}{16 T^2}\frac{s_1+s_2}{\frac{\alpha(0)}{T}
\left(s_1+s_2\right)+s_1s_2},
\label{gam-m}
\end{equation}
and
\begin{equation}
\alpha\left(\vec q \right)=\frac{1}{4}\left[ E-T\ln 2+
\sum_{\vec R}V\left(\vec R\right)\exp(-i\vec q \vec R)\right].
\label{alpha}
\end{equation}
where the summation in Eqn.~(\ref{gam-m}) is carried out over the
first Brillouin zone (BZ).
The next step would be the evaluation of the partition function, however,
instead of doing it we proceed as in the study of the CSM. That is,
we find
the saddle points of $\Z$ and study its temperature behavior since it
contains the complete information of the possible phase transitions of
the system. Nonetheless, it is important to note that the
evaluation of the partition function by the steepest descendent method
becomes exact in the thermodynamic limit $\N \rightarrow \infty$, as in 
the CSM.
The saddle points of the integrand of $\Z$ are determined by the
following equations:
\begin{eqnarray}
\frac{\partial\gamma}{\partial s_1}=0&=&1-\frac{1}{2(s_1+s_2)}
-\frac{1}{2\N}\frac{s_2^2}{s_1+s_2}\sum_{\vec q}
\frac{1}{s_1s_2+\frac{\alpha(\vec q)}{T}\left(s_1+s_2\right)}\nonumber\\
&-&\frac{H^2}{16T^2}\frac{s_2^2}{\left(\frac{\alpha(0)}{T}
\left(s_1+s_2\right) +s_1s_2\right)^2},
\label{saddle}
\end{eqnarray}

\begin{eqnarray}
\frac{\partial\gamma}{\partial s_2}=0
&=&1-\frac{1}{2(s_1+s_2)}
-\frac{1}{2\N}\frac{s_1^2}{s_1+s_2}\sum_{\vec q}
\frac{1}{s_1s_2+\frac{\alpha(\vec q)}{T}\left(s_1+s_2\right)}\nonumber\\
&-&\frac{H^2}{16T^2}\frac{s_1^2}{\left(\frac{\alpha(0)}{T}\left(s_1+s_2\right)
+s_1s_2\right)^2}.
\label{saddle2}
\end{eqnarray}
These equations are symmetric in the variables $s_1$ and $s_2$.
Substracting  Eqn.~(\ref{saddle2}) from Eqn.~(\ref{saddle}) and
taking into account
that $\gamma(s_1,s_2)$ has to be an analytic function at the  saddle point,
one gets the condition $s_1=s_2=s$. Substituting this in either, 
Eqns.~(\ref{saddle}) or (\ref{saddle2}),
we obtain the equation that yields the saddle points of the integrand of
the partition function, namely
\begin{equation}
\psi (s)=\frac{1}{\N}\sum_{\vec q}\frac{1}{s+2\frac{\alpha(\vec q)}
{T}}+\frac{H^2}{4T^2}\frac{1}{\left(s+2\frac{\alpha(0)}{T}\right)^2}+
\frac{1}{s}=4.
\label{ss}
\end{equation}
To investigate the existence of phase transitions in the model, we should
recall that, as in the CSM, the existence of a saddle point is associated
with the setting in of a disordered phase, PM phase in this case,
while the absence of a saddle point is related to the realization of an
ordered phase, FM or AFM. The critical temperature is obtained 
by studying the evolution of the saddle point as a function of temperature.
To be more precise, the phase transition occurs just when
the saddle point dissapears. As we will see below there is a such transition
in some region of the parameters space $(E,J,H)$, nonetheless, there is another
region in the parameters space  where a saddle point does exist for any
temperature, indicating the absence of a phase transition in such region.
For instance, one can easily show that for a nonzero external 
magnetic field and if the ferromagnetic interactions between spins
satisfy the condition $\min_{\>\vec q}\alpha(\vec q)=\alpha(0)$, then
a saddle point exists in the whole parameter space and therefore no phase
transition takes place.
However, for antiferromagnetic interactions one gets 
$\min_{\>\vec q}\alpha(\vec q)<\alpha(0)$ and a more
careful and detailed  analysis should be carried out to obtain the 
corresponding phase diagram.(See subsection \ref{nonzeroh}).
In the following section we apply our general model hamiltonian to 
study a system with short range interactions.        

\section{\bf Phase transitions in a ferromagnetic and an antiferromagnetic
systems with short-range interactions.}
\label{short-range}
\subsection{ Zero external magnetic field.}
\label{zeroh}

Let us consider a magnetic system with short range interactions,
that is, $V(\vec R-\vec R')=-J$ for nearest neighbors
$\vec R,\vec R'$, and zero otherwise.
In this case Eqn.~(\ref{alpha}) becomes:
\begin{equation}
\alpha(\vec q)=\frac{1}{4}\left(E-T\ln 2-
2J\sum_{j=1}^{d} \cos q_j\right),
\label{alpha1}
\end{equation}
where $d$ is the spatial dimensionality, $\vec q=(a_1 q_1,a_2 q_2,
.....)$ is a dimensionless vector in the reciprocal space and
$a_j$ is the lattice constant
along the $j$ direction.
Setting $H=0$ in Eqn.~(\ref{ss}), the function 
$\psi\left(s\right)=\psi_{0}\left(s\right)$
can be rewritten as
\begin{equation}
4=\psi_{0}\left(s\right)=\frac{1}{s}+\frac{T}{J}\frac{1}{(2\pi)^d}
\int_{B.Z.} d\vec q\frac{1}{\frac{T}{J}\left(s+\frac{E}{2T}-
\frac{1}{2}\ln 2\right)-\left(\sum_{j=1}^{d} \cos q_j\right)},
\label{sss}
\end{equation}
here the integration is carried out in the first Brillouin zone, that is,
in the range  $-\pi<q_j<\pi$ for each $j$.
As we know and according to the definition of the interaction potential,
$J>0$, for ferromagnetic interactions while
$J<0$, for antiferromagnetic interactions. Notice that
in the framework of our model the phase diagram in the $E$ versus $J$ plane
for the antiferromagnet ($J<0$) is obtained from the reflection about the
line $J=0$ of the phase diagram
of the ferromagnet. Indeed, by introducing a shift in the wave vectors in
$q$-space, that is,  by replacing $\vec q$ by 
$\vec q^{\>*}=(q_1-\pi,q_2-\pi, ...)$
one can show that $J\sum \cos q_j=-J\sum \cos q^{\>*}_j$. Therefore,
Eqn.~(\ref{sss}) for the antiferromagnetic system results in that
corresponding to a ferromagnet. The only difference being
the critical value $\vec q=\vec q_{\rm m}$ at which $\alpha (\vec q)$ 
attains its minimum value. For a ferromagnet this minimum is reached
at $\vec q_{\rm m}=0$ while for an antiferromagnet the minimum occurs at
$\vec q_{\rm m}=(-\pi,-\pi,...)$.
In term of the orientations of spins this is equivalent to say that
nearest-neighbors spins try to align parallel for ferromagnetic interactions
or anti-parallel for antiferromagnetic interactions. Thus, all particularities
of the phase transition of the antiferromagnet are the same as those for 
the ferromagnet. Bearing this in mind we carry out a detailed
analysis of the phase diagram for a ferromagnet
and extend directly the results for an antiferromagnet by carrying out a
reflection about $J=0$.

The next step is to examine the properties of $\psi_{0}(s)$ for
a ferromagnet since they determine the phase transitions in the 
model under investigation.
It is easy to see that $\psi_{0}(s)$ decreases monotonically 
and is analytic in the interval
$s_0(T) < s \leq +\infty$, where
$s_0$ is defined as,
\begin{equation}
s_0=\max\left(0, \frac{1}{2}\ln 2 -(E-2dJ)/2T\right).
\label{s0}
\end{equation}
The behavior of the function $\psi_{0}(s)$ is such that it can take
values from zero when $s\rightarrow\infty$ 
up to a nonzero value $\psi_{0}(s_0)$.
Its  behavior is shown in Fig. \ref{fig-psi-s}, where we have indicated
the positions of $s_0$ (points {\bf B} and {\bf A}) at two temperatures 
$T_{1}$ and $T_{2}$ that are below and above the critical temperature $T_{c}$,
respectively. The point {\bf R} represents the saddle point which existence
is related to the set in of the disordered phase.
Thus, at $T=T_{2}$ (point {\bf P} and dot dashed line), $\psi_{0}(s_0)>4$, 
and Eqn.~( \ref{sss}) has a root, indicative of the existence of a saddle
point and a paramagnetic phase. In the opposite case, at
$T=T_{1}$ (point {\bf Q} and solid line), $\psi_{0}(s_0)<4$, and
Eqn.~( \ref{sss} ) has no root yielding no phase transition to the 
paramagnetic phase. Hence it is necesary to analyze the behavior of
$\psi_{0}(s_0)$ as a function of temperature in the parameters space $(E, J)$
to be able to construct the phase diagram.
First of all, we have to study the general conditions for
which the inequality $\psi_{0}(s_0)<4$,
responsible for the ferromagnetic phase, is valid.
Thus, in the region of parameters $(E$, $J, T)$
where $s_0=0$, the function $\psi_{0}(s_0)\rightarrow\infty$
($\psi_{0}(s_0)>4$)  and a paramagnetic phase exists.
Nonetheless, when $s_0>0$ it is necessary to analyse the convergence
of the  integral for $s=s_0$  in Eqn.~(\ref{sss}).
In doing so we substitute $s=s_0$ and expand $\cos q_j\approx 1-q_j^2/2$,
yielding,
\begin{equation}
\int_{B.Z.} d\vec q\frac{1}{\frac{T}{J}\left(s_0+\frac{E}{2T}-
\frac{1}{2}\ln 2\right)-\left(\sum_{j=1}^{d} \cos q_j\right)}\sim
\int_{B.Z.} \frac{q^{d-1}dq}{q^2},
\label{integ}
\end{equation}
where we have omited an irrelevant constant numerical factor.
The last integral converges if the spatial dimension $d$ is such that,
$d-3>-1$, i.e. for spatial dimensionality  $d>2$.
This means that for $d \leq 2$, $\psi_{0}(s\rightarrow s_0)\rightarrow \infty$
yielding a saddle point
for any values of $E,J$ and $T$, and no phase transition
takes place for lower dimensional systems.
A more detailed analysis has to be carried out for three dimensions because
the integral (\ref{integ}) does converge and
can be evaluated as \cite{berlin-kac},
$${{1}\over{(2\pi)^3}}\int \frac{d\vec q}{3-\sum_{j=1}^{3}
\cos q_j} \approx 0.50541 = I_{\rm cr}.$$
Introducing this result in Eqn.~(\ref{sss}) one obtains, for $s_0(T)>0$,
\begin{equation}
\psi_{0}(s_0(T), T)=\frac{T}{J}I_{\rm cr}+\frac{2T}{T\ln 2-E+6J}.
\label{psi0}
\end{equation}
In this case we find (see Fig. \ref{fig-psi0-1}) that there are
regions in the parameter space $(E,J,T)$  where $\psi_{0}(s_0)<4$, i.e. 
the saddle point
disappears. This happens for temperatures below a certain critical 
temperature, $T_{c}$, indicating the transition to the
ferromagnetic (antiferromagnetic) state.
A more detailed analysis of $\psi_{0} \Big( s_{0}(T),T \Big)$ shows 
that  there are, actually, two ranges of
parameters to be considered: (1) $E-6J<0$ in which case $s_{0}(T)$
goes from $+\infty$ at
$T=0$ up to $\frac{1}{2}\ln 2$ at $T=\infty$, and (2) $E-6J>0$
for which $s_{0}(T)$ goes
from zero at $T=0$ up to $\frac{1}{2}\ln 2$ at $T=\infty$.

Let us first consider the behavior of $\psi_{0}(s_0(T),T)$ as a function
of temperature when $E-6J<0$. In this domain of parameters $s_0>0$ for any $T$
and Eqn.~(\ref{psi0}) is always valid.
At very low temperatures  $\psi_{0}(s_0(T),T)<4$ and at $T=0$, $\psi_{0}(s_0)$
goes down to zero, implying that 
there is no saddle point, thus
the system settles down in a ferromagnetic state.
At higher temperatures, the function $\psi_{0}(s_0(T),T)$
increases almost linearly with temperature and therefore
$\psi_{0}(s_0)$ crosses over the line $\psi_{0}(s_0)=4$, 
at the critical temperature and the saddle point shows up, hence the
system goes from the ferromagnetic (antiferromagnetic) phase to the
paramagnetic one.
Thus, this region of parameters yields one critical temperature that
corresponds to the  usual PM-FM (or PM-AFM) phase transition.
On the other hand, in the range of parameters where $E-6J>0$ the behavior of
$\psi_{0}\big( s_0(T),T\big)$ is different. For  $T<(E-6J)/\ln 2$, the function
$\psi_{0}(s_0)$ diverges since $s_0=0$. For $s_0(T)>0$, 
the function $\psi_{0}\big(s_0(T)\big)$
decreases asymptotically and quite rapidly  from $\infty$ at 
$T=(E-6J)/\ln 2$ until
it reaches a minimum at the temperature $T^*=T^*(E,J)$, given by
\begin{equation}
T^{*}=\frac{1}{\ln 2} \Big( E-6J+\sqrt{\frac{2J(E-6J)}{I_{\rm cr} } }
\>\> \Big),
\label{tstar}
\end{equation}
\noindent
and afterwards, it begins to increase. This behavior is shown
in detail in Fig. \ref{fig-psi0-1}. 
It is obvious that as soon as $\psi_{0}(s_0(T^*),T^*)>4$ then
$\psi_{0}(s_0(T),T)>4$ for any temperature $T$ greater than $T^{*}$ and the
paramagnetic phase is only present, therefore,
no phase transition occurs.
This is what happens when $\frac{E}{J} > 6+x_{c}$ for any value of $T$,
where, the value of $x_{c}$ is given by
$x_{c}=(1/I_{cr})(2\sqrt{\ln 2} - \sqrt{2})^{2}\approx 0.124 \>$. 
The corresponding curves
$\psi_{0}(s_0(T),T)$ that fullfil this condition are shown in
Fig. \ref{fig-psi0-1}.
However, if $6< \frac{E}{J} < 6+x_{c}$ then
$\psi_{0}(s_0(T^*),T^*)<4$, and $\psi_{0}(s_0(T),T)$ intersects
the line $\psi=4$ in two points, yielding two critical temperatures:
$T_{c}^{+}$, the branch where the PM-FM (PM-AFM) phase transition occurs,
and $T_{c}^{-}$, the branch at which the system goes back to the paramagnetic
phase; so a reentrant phase transition takes place. These critical temperatures
are given by
\begin{equation}
T^{\pm}_{c}=\frac{(E-6J)I_{\rm cr}+4J\ln 2-2J \pm \D}
{2I_{\rm cr} \ln 2},
\label{Tc}
\end{equation}
with
\begin{equation}
\D=\Big[ \big( (E-6J)I_{\rm cr}+4J\ln 2-2J\big)^2
-16(E-6J)I_{\rm cr} J\ln 2 \Big]^{1/2}.
\end{equation}
From these results we can state that
in the temperature range $T_{c}^{-}\> < T < \> T_{c}^{+}$
a magnetic ordered state takes place,
whereas  outside of this interval, i.e. at lower  $T<T_{c}^{-}$
and higher
$T>T_{c}^{+}$ temperatures, a paramagnetic state settles in. Translating
these results in the parameter domain we can say that,
the region $6< \frac{E}{J} < 6+x_{c}$  where the reentrant phase transition
occurs, separates out the region $E-6J<0$ where the usual PM-FM (PM-AFM) phase
transition happens,  from the region $6+x_{c}<\frac{E}{J}$  where there is
no  phase transition at all. 

To make more intuitive and explicit the above results  we
have plotted the two dimesional phase diagram
$\frac{T^{\pm}_{c}}{J}$ versus $\frac{E}{J}$ in figure
\ref{ph-diag1-2d}. The solid line represents the
critical temperature $T^{+}_{c}$ that corresponds to the usual high 
temperature single phase
transition while the dashed line represents the novel and intriguing low
temperature reentrant phase transition $T^{-}_{c}$. These lines have also been
plotted in a larger scale in the inset, where the two branches are clearly seen.
The two critical temperatures become gradually closer to each other
with increasing $E/J$ until the ordered phase dissapears completely.

\subsection{ Nonzero uniform magnetic field.}
\label{nonzeroh}

In this subsection we investigate how the phase diagram of the generalized
BEG model changes in the presence of an homogeneous magnetic field
in three spatial dimensions.
In the case of ferromagnetic interactions the application of an
external magnetic field terminates the transition to the state with
a spontaneous magnetization since a
saddle point appears for any values of $E$, $J$ and $T$, as it just happens
for the CSM\cite{berlin-kac}. That is, the external magnetic field is a
symmetry breaking field that produces a magnetized state that always yields
a saddle point in the integrand of the partition function for any values 
of the parameters. Because of this,  an additional analysis should be done
to study the thermodynamics of the present model,
for instance, one can study the behavior of the magnetic susceptibility, as
was done in \cite{berlin-kac} for the CSM. 
For the time being, it is the purpose
of the present paper to study the phase diagram and leave the study of the
thermodynamics for a forthcoming paper\cite{thermo00}.
Let us now focus our attention on the case
of antiferromagnetic interactions. Basically, we will study the circumstances
under which the saddle point dissapears in the presence of an homogeneous
magnetic field, i.e. when the  antiferromagnetic ordering sets in.
Using Eqn.~(\ref{ss}) and the results of section \ref{zeroh} it is easy to
show that $\psi(s)$ is an analytic function that decreases monotonically
in the interval $s_0<s<\infty$, where $s_0$ is given by Eqn.~(\ref{s0}).
To be able to use the results obtained for an antiferromagnet in a 
zero magnetic field, in what follows we will denote the modulus of the
AFM exchange interaction as $J>0$ since the minus sign has already been
included as a shift in the reciprocal space vectors, that is, we have 
replaced $\vec q$ by $\vec q\>^{*}$.
Henceforth, in analogy  with the case of zero magnetic field,  whenever
$\psi(s_0(T),T,H)>4$, a saddle point does exist and it
yields a paramagnetic phase, whereas in the
opposite case, $\psi(s_0(T),T,H)<4$, the system  
settles in an antiferromagnetic state, 
where the function $\psi(s_0(T),T,H)$ is given by
\begin{equation}
\psi(s_0(T),T,H)= \psi_{0}(s_0(T),T) +\Big( \frac{H}{12J}\Big)^2 = 
\frac{T}{J}I_{\rm cr}+ \frac{2T}{T\ln 2-E+6J} 
+\Big( \frac{H}{12J}\Big)^2,
\label{psi0H}
\end{equation}
To obtain this equation we have made the following substitution in
Eqn.~(\ref{ss}),
$$
H^2/(4T^2(s_0+2\alpha(\vec q=0)/T)^2)=\Big( \frac{H}{12J}\Big)^2.
$$
As in the previous subsection, the equation $\psi_{0}(s_0(T_c),T_c,H)=4$
determines the
temperature at which the phase transition occurs. From this very last
condition one can rewrite Eqn.~(\ref{psi0H}) as ,
\begin{equation}
\psi_{0}(s_0(T_c),T_c)=4-\Big( \frac{H}{12J}\Big)^2,
\end{equation}
where $\psi_{0}(s_0(T),T)$ is the function that we have carefully studied
for the ferromagnetic and antiferromagnetic systems in abscence of 
an external field.
Thus, one obtains $T_{c}(H)$ by looking at the intercept of the curves
$\psi_{0}(s_0)$ shown in Fig. \ref{fig-psi0-1}, with the lines $4-(H/12J)^2$.
The resulting plots are shown in Fig. \ref{fig-psi0-H}
for the following values of the reduced
magnetic field, $\frac{H}{J}$: (a) 2, (b) 8, (c) 16 and (d) 20.
One  immediately sees that in the parameter
region $E-6J<0$, there is only one root  of the equation
$\psi(s_0(T_c),T_c,H)=4$,
what signals the usual transition to the antiferromagnetic phase.
The critical temperature of this transition,
$T_{c}(H)$, that is now a function of the external
magnetic field, diminishes as the field intensity increases and becomes equal
to zero at the critical field value $H_{c2}=24J$. Thus, the phase transition
dissapears for fields such that $H>H_{c2}$,
and the paramagnetic state settles in at all temperatures, as
one naturally would expect, since the antiferromagnetic ordering is
broken by higher external field intensities. We have represented this
critical field value by $H_{c2}$ because of, as we will show below, there is
another critical field value $H_{c1}\approx 0.5H_{c2}$ at which the reentrant
phase transition dissapears.

On the other hand,
in the parameter region where $E-6J>0$ there are two possibilities:
(i) the reentrant AFM-PM phase transition or
(ii) the absence of a phase transition at all.
The domain of parameters where the
reentrant transition takes place is restricted by the inequalities,
$E-6J>0$ and $\psi_{0}(s_0(T^*),T^*)<4-(H/12J)^2$, where $T^*$ is defined
in Eqn.~(\ref{tstar}). These two relations can be reexpressed in the form:
$6<E/J<6+x_{c}(H)$, with $x_{c}(H)$ given by
\begin{equation}
x_c = \frac{
\left(2\sqrt{\ln 2}\sqrt{1-(\frac{H}{H_{c1}})^2}-\sqrt{2}\right)^2}
{I_{\rm cr} }.
\end{equation}
The two critical temperature branches of the phase transitions 
are the roots of the equation $\psi(s_0(T_c),T_c,H)=4$, and are given by
\begin{equation}
T_c^\pm=\frac{(E-6J)I_{\rm cr}-2J+J\Big(4-(H/12J)^2\Big)\ln 2 \pm \D(H)}
{2I_{\rm cr}\ln 2},
\end{equation}
where
\begin{equation}
\D=\sqrt{\left((6J-E)I_{\rm cr}+2J-J\left(4-\Big(\frac{H}{12J}\Big)^{2}
\right)\ln 2\right)
+4I_{\rm cr}J(6J-E)\left(4-(\frac{H}{12J})^2\right)\ln 2}.
\end{equation}
Evidently, the reentrance phenomenon is suppressed by the external
field  when it exceeds the critical value $H_{c1}$ defined as the
root of the equation $x_c(H_{c1})=0$.
This leads to the critical field value,
\begin{equation}
H_{c1}=H_{c2}\frac{\sqrt{4-2/\ln 2}}{2}=0.52788H_{c2}\approx \frac{H_{c2}}{2}.
\end{equation}
These results allow us to construct the phase diagram $(\frac{T}{J}$ versus
$\frac{E}{J},\frac{H}{J})$ which, for the sake of clarity, 
is shown in two parts, Figs. \ref{ph-diag2a-3d}
and \ref{ph-diag2b-3d}. The former shows the region in three dimensional
space where there is a single transition, while the latter corresponds to
the region where the reentrance phenomenon takes place. That is, the
whole half-plane $(\frac{E}{J},\frac{H}{J})$
splits into three regions, the first one corresponds to the
{\it ``single transition"} characterized by the PM-AFM phase transition, while
the {\it ``reentrant AFM-PM phase transition"} occurs in the second region.
The third region corresponds to the domain where there is no phase transition
at any temperature and any values of the coupling parameters $E$ and $J$.
\section{Exact global spherical condition.}
\label{sec-global}
One of the main shortcomings of SM approach, that we have used here,
Eqn.~(\ref{spher1}),
is that it  fails when spatial and spin dimensionality are low.
For instance, it is known that in two dimensions the CSM approximation for an
Ising ferromagnet does not describe the correct transition scenario.
Nonetheless, for spin dimensions greater than one,
the CSM approximation works quite well. In fact, it is
exact in the limit of infinite spin dimensionality\cite{stanley68}.
The main drawback of the classical spherical condition:
$\sum_{\vec R}(a^2_{\vec R}-1)=0$  is that it holds not only for ``right 
configurations", $a_{\vec R}=\pm 1$, but it also does for the ``wrong
configurations" where some of the spin variables are $a_{\vec R}<1$ while 
others are  $a_{\vec R}>1$.

So, in trying to fix up this shortcoming we replace the usual
spherical condition by the ``general global" one, namely
\begin{equation}
\sum_{\vec R}(a^2_{\vec R}-1)^2=0.
\label{new-cond}
\end{equation}
This sum is equal to zero only for the ``right configurations"
$a_{\vec R}=\pm 1$. Therefore,  the model with continuously
varying spin variables plus the ``global spherical condition",
Eqn.~ (\ref{new-cond}), are
equivalent to the original spin one BEG hamiltonian.
Note that we may also use the general global condition
$\sum_{\vec R}f(a^2_{\vec R}-1)=0$ where $f(x)>0$ for $x\ne 0$
and $f(0)=0$, however, for definiteness and simplicity
one can use Eqn.~(\ref{new-cond}).
This condition can be expressed in a form slightly closer to the usual
spherical condition as follows:
\begin{equation}
\N=\sum_{\vec R}a^2_{\vec R}+\sum_{\vec R}(a^2_{\vec R}-1)(a_{\vec R}^2-2).
\end{equation}

With this ``new global condition"  we can write a new effective hamiltonian
which in turn  defines the partition function $\tilde\Z(s_1,s_2)$ (see,
Eqns.~(\ref{tz}) and (\ref{th})):
\begin{equation}
{\tilde\H}(s_{1},s_{2})={\tilde\H}_0+
Ts_1\sum_{\vec R}(a_{\vec R}^2-1)(a_{\vec R}^2-2)+
Ts_2\sum_{\vec R}(b_{\vec R}^2-1)(b_{\vec R}^2-2),
\end{equation}
where $\tilde\H_0$ is the usual spherical effective hamiltonian defined by
Eqn.~(\ref{th}). So one can rewrite the new effective hamiltonian
as the usual spherical hamiltonian plus an additional interaction
term $U(s_{1},s_{2})$, that is: ${\tilde\H}(s_{1},s_{2})=
{\tilde\H}_0+U(s_{1},s_{2})$.
Note that this new effective hamiltonian is much more complicated than
the one we have studied in this paper and,
unfortunately, the interaction $U(s_{1},s_{2})$ cannot be  
considered perturbatively because it lacks of a
small parameter. However, our propossal allows us, in principle,
to improve the classical spherical model for a system with lower spin and space
dimensionalities. The problem of incorporating the $U(s)$  interaction
into the complete solution is a difficult task and it is the subject of
current research.

\section{\bf Summary and conclusions}
\label{conclusions}
In this paper we have introduced and studied the phase diagrams of a 
generalized spherical version of the BEG model considering ferromagnetic
as well as antiferromagnetic interactions.
The model hamiltonian involves terms representative of quadrupole interactions
and  magnetic spin interactions in zero and nonzero external magnetic field.
We have shown explicitly that in the short range interactions limit
and zero external magnetic field the model presents no phase transition
in one and two spatial dimensions. However, in three dimensions
it undergoes a regular PM-FM(AFM) transition
at nonzero temperature. This happens in the case when the magnetic interactions
dominate over the quadrupole  ones. Nonetheless, in
the range of parameters where quadrupole  and magnetic
interactions are relevant we obtain a low temperature
reentrant FM(AFM)-PM phase  transition, in
addition to the conventional PM-FM(AFM)
phase transition at higher temperatures. These phase transitions are absent
in the strong quadrupole  interaction regime.
On the other hand, when there is an external magnetic field there is 
no phase transition for the ferromagnet in three dimensions,
since there appears a permanent magnetized state that terminates
the  PM-FM transition. 
This result appears to be similar to the one obtained in the 
CSM\cite{berlin-kac}, and it is a motivation to further 
investigate the thermodynamics of the present model\cite{thermo00}.
However, in the case of antiferromagnetic interactions and nonzero magnetic
field we also get the reentance phenomenon whenever the magnetic spin and
quadrupole interactions compete between them and always that 
$H<H_{c1}\approx 12J$. If 
$H_{c1} \leq H$ the reentrant phase transition dissapears.
We also find a second critical value of the external magnetic field $H_{c2}$,
above which the PM-AFM phase transition dissapears.
Finally, we have introduced a ``novel global spherical condition"  that
makes our approach valid even for low spin dimensionality. However, 
this led us to a more complicated
hamiltonian which phase diagram is under current investigation.
\section{\bf Acknowledgements}
This work has been supported by CONACYT-MEXICO grant No. 25298-E.

\newpage
%{\Large \bf Figure Captions}

\begin{center}
\begin{figure}
\caption{ 
Behavior of $\psi$ as function of the parameter
$s$ at two temperatures $T_{1}$ and $T_{2}$ below and above
the critical temperature, respectively. Points {\bf A} and {\bf B}
have coordinates $\Big( s_{0}(T_{2}), 0\Big)$ and $\Big( s_{0}(T_{1}), 0\Big)$,
while the coordinates of points {\bf P} and {\bf Q} are 
$\Big( s_{0}(T_{2}), \psi \Big)$  and $\Big( s_{0}(T_{1}), \psi \Big)$,
respectively. Point {\bf R} indicates the saddle point that signals
the phase transition.
Note that  $\psi(s)$ decreases
monotonically with $s$, that is, it goes from $\psi(s_0)$ and goes down
to zero asymptotically as $s\rightarrow \infty $.
}
\label{fig-psi-s}
\end{figure}
\end{center}

\begin{center}
\begin{figure}
\caption{ 
Behavior of $\psi_{0} (s_0)$ as function of $\frac{T}{J}$
and different values of $\frac{E}{J}$. Note that for $0< \frac{E}{J}<6$
there is only one intercept of these curves with the line $\psi_{0}=4$,
as an indication of a single transition. 
However, for $6 \leq \frac{E}{J}<6+x_{c}$
we see a lower temperature intercept corresponding to the reentrant
phase transition.
}
\label{fig-psi0-1}
\end{figure}
\end{center}

\begin{center}
\begin{figure}
\caption{
Phase diagram for the ferromagnet where the
reentrant transition appears at $\frac{E}{J} > 6$. The letters {\bf P}
and {\bf F} indicate the paramagnetic and ferromagnetic states,
respectively. In the inset
we show the part of the phase diagram corresponding to the
reentrance behavior in the parameter region $6<\frac{E}{J} < 6+x_{c}$.
}
\label{ph-diag1-2d}
\end{figure}
\end{center}

\begin{center}
\begin{figure}
\caption{ 
Behavior of $\psi_{0}(s_0)$ versus $\frac{T}{J}$ for
different values of the parameter $\frac{E}{J}$ (same values  as in Fig.
\ref{fig-psi0-1})
and different values of the external magnetic field.(a) $\frac{H}{J}=2$
(b) $\frac{H}{J}= 8$, (c) $\frac{H}{J}=16$, and (d) $\frac{H}{J}= 20$.
Notice that for $E/J<6$, where a single transition occurs, 
the critical temperatures $T_{c}(H)$,
decrease as the magnetic field increases. On the other hand,
for curves $\psi (s)$, with $6 < \frac{E}{J} < 6+x_{c}(H)$,
there are two transitions,
the high temperature one and the one corresponding to the reentrance to the
paramagnetic state, panels (a) and (b),
while in panels (c) and (d) there is no phase transition at all, because of the
fact that $\frac{H}{J} > \frac{H_{c}}{J}$.
}
\label{fig-psi0-H}
\end{figure}
\end{center}

\begin{center}
\begin{figure}
\caption{
Three dimensional phase diagram $\frac{T_c}{J}$ versus
$\frac{E}{J},\frac{H}{J}$ in the range of parameters where there is a single
transition.
}
\label{ph-diag2a-3d}
\end{figure}
\end{center}

\begin{center}
\begin{figure}
\caption{
Three dimensional phase diagram $\frac{T_c}{J}$ versus
$\frac{E}{J},\frac{H}{J}$ restricted to the range of
parameters where the reentrant transition takes place.
}
\label{ph-diag2b-3d}
\end{figure}
\end{center}
\end{document}